\documentclass[journal]{IEEEtran}
\usepackage{amsmath,amsfonts,amssymb}
\usepackage{algorithmic}
\usepackage{algorithm}
\usepackage{array}
\usepackage[caption=false,font=normalsize,labelfont=sf,textfont=sf]{subfig}
\usepackage{textcomp}
\usepackage{stfloats}
\usepackage{url}
\usepackage{verbatim}
\usepackage{cite}
\usepackage[dvipsnames]{xcolor}
\usepackage[normalem]{ulem}
\usepackage{mathtools}
\ifCLASSINFOpdf
   \usepackage[pdftex]{graphicx}
\else
   \usepackage[dvips]{graphicx}
   \DeclareGraphicsExtensions{.eps}
\fi

\graphicspath{{./figures/}}

\DeclareMathOperator{\sinc}{sinc}
\newcommand{\red}[1]{\textcolor{Red}{#1}}

\DeclareRobustCommand{\erase}{\bgroup\markoverwith{\textcolor{red}{\rule[.5ex]{2pt}{0.5pt}}}\ULon}
\newcommand{\del}[1]{\red{\erase{#1}}}
\newcommand{\add}[1]{\textcolor{blue}{#1}}

\begin{document}

\title{Error Performance of Rectangular Pulse-shaped OTFS with Practical Receivers}

\begin{comment}
\author{\IEEEauthorblockN{Cheng Shen, Jinhong Yuan}
\IEEEauthorblockA{The University of New South Wales \\
Sydney, NSW, 2052, Australia \\
cheng.shen1@unswalumni.com, j.yuan@unsw.edu.au
\and
\IEEEauthorblockN{Hai Lin}
\IEEEauthorblockA{Osaka Prefecture University \\
Sakai, Osaka, 599-8531, Japan \\
hai.lin@ieee.org}
}}
\end{comment}
\author{Cheng Shen, Jinhong Yuan~\IEEEmembership{Fellow,~IEEE}, and Hai Lin~\IEEEmembership{Senior Member,~IEEE}
        % <-this % stops a space
%\thanks{C. Shen and J. Yuan are with the School of Electrical Engineering and Telecommunications, University of New South Wales, Sydney, NSW 2052,
%Australia (e-mail: z5141506@zmail.unsw.edu.au; j.yuan@unsw.edu.au) H. Lin is with ...}% <-this % stops a space
\thanks{Manuscript received April 1, 2022; revised August 16, 2022.}}

\markboth{Journal of \LaTeX\ Class Files,~Vol.~14, No.~8, August~2021}%
{Shell \MakeLowercase{\textit{et al.}}: A Sample Article Using IEEEtran.cls for IEEE Journals}


\maketitle

\begin{abstract}
In this letter, we investigate error performance of rectangular pulse-shaped orthogonal time frequency space (OTFS) modulation with a practical receiver. Specifically, we consider an essential bandpass filter at receiver RF front-end, which has been ignored in existing works. We analyse the effect of rectangular pulses on practical OTFS receiver performance, and derive the exact forms of interference in delay-Doppler (DD) domain. We demonstrate that the transmitted information symbols in certain regions of the DD domain are severely contaminated. As a result, there is an error floor in the receiver error performance, which needs to be addressed for such OTFS waveform in practical systems.
\end{abstract}

\section{Introduction}

\IEEEPARstart{A}{s} a 2D modulation technique, orthogonal time frequency space (OTFS) has recently been proposed to combat nasty doubly-selective channels in future communication systems\cite{Hadaniwcnc17,hadani2018arxiv}. In OTFS, information is modulated in delay-Doppler (DD) domain and thus coupled with a relatively stable and compact DD domain channel model, which facilitates a simple channel estimation and accurate detection for time-varying channels. This leads to superior performance of OTFS over OFDM when channels are characterised with increasingly larger Doppler spread. %, due to higher mobility and higher carrier frequencies in emerging applications. 
It is shown in \cite{Hadaniwcnc17} that if the OTFS transmitter and receiver pulses satisfy the \emph{bi-orthogonality robustness condition} with respect to the time-frequency grid over the support of the channel response, OTFS signals can be detected with a simple one-tap equalizer. However. such pulses are not realizable in practice\cite{Hadaniwcnc17} \cite{viterbo_tvt_19}. Alternatively, %\del{one of} 
one of the most widely investigated OTFS signals was introduced in \cite{viterbo_tvt_19}. It employs a rectangular pulse at the transmitter and receiver to form a signal structure identical to OFDM without slot-wise cyclic prefix (CP), but only adds a single frame-wise CP at the start of an OTFS frame.
%and adds a single cyclic prefix (CP) at the start of an OTFS frame. 
Compared to CP-based OTFS, which employs a CP for each time slot of the OTFS frame\footnote{The CP-based OTFS can be interpreted as pre-processed OFDM and implemented as an overlay to existing OFDM systems ~\cite{Hadaniwcnc17,hadani2018arxiv}.}, this reduced cyclic prefix (RCP) OTFS scheme has a lower overhead and therefore a higher spectral efficiency. 
%\del{maintains a sparse channel input-output relation.}  
It has become a focal point in OTFS research community, especially for those investigating its practical applications.

A major drawback of rectangular pulse is its high out-of-band emission (OOBE), because the rectangular pulse belongs to the family of time-limited pulses and has a slow-decaying sinc shape in frequency domain.
For any practical OFDM systems, OOBE needs to be carefully handled to avoid interference to adjacent channels or self-interference in a band-limited transmission medium, where the time-limited pulses can lead to severe performance degradation \cite{RChang1966}.

Since the RCP-OTFS relies on the rectangular pulse based CP-free OFDM through an inverse simplectic finite Fourier transform (ISFFT) precoder, OOBE is also an inevitable practical issue for RCP-OTFS~\cite{oddm}. However, the ISFFT precoder prevents the application of the conventional OOBE suppression approaches adopted in OFDM to the RCP-OTFS. For example, spectral
guarding or frequency-domain precoding based OOBE mitigation methods will break the inherent connection between the DD and time-freqeuncy (TF) domain signals governed by the ISFFT precoder.
%is not easy to control the spectrum of OTFS signal, given the DD domain modulation nature of OTFS.
Meanwhile, the absence of CP and cyclic suffix (CS) for each time slot makes windowing based OOBE mitigation methods infeasible, while replacing the rectangular pulse with a spectrally compact pulse leads to severe performance degradation \cite{viterbo_tvt_19}, due to loss of orthogonality.

% \red{Addressing OOBE for RCP-OTFS with rectangular pulse is also an important issue~\cite{oddmicc}. However, this topic has not been investigated much in the literature. While OTFS shares some similarities with OFDM, conventional OOBE reduction methods for OFDM systems cannot be employed directly in such RCP-OTFS frame structure.} 
% %\add{How to mitigate OOBE for RCP-OTFS seems not to be a trivial task. should we add this sentence?} 
% \red{For example, the absence of CP for each time slot makes windowing based OOBE mitigation methods infeasible, as no redundant time exists at each time slot for smoothing the sharp edges of rectangular windows. In addition, directly replacing the rectangular window with a spectrally compact window can lead to significant performance degradation [3], due to loss of orthogonality between subcarriers. Furthermore,  spectral
% guarding or frequency-domain precoding based OOBE mitigation methods will not work either, as it is not easy to control the spectrum of OTFS signal, given the DD domain modulation nature of OTFS.} 

In spite of the importance of OOBE in practical systems, OOBE for RCP-OTFS has not been investigated much in the literature. In particular, existing works on RCP-OTFS assume that without suppressing the OOBE, the transmitted RCP-OTFS waveform is corrupted by the {channel dispersion} and noise only. However, even the RCP-OTFS signal with high OOBE is allowed to be emitted into the air, the receiver always needs to filter the received signal first. A bandpass filter (BPF) not only prevents the interference and noise from other wireless channels, but also ensures that no aliasing occurs in the subsequent sampling. In other words, a BPF at the receiver's RF front-end is a \emph{must} before all sampling and baseband digital processing operations. Unfortunately, such a vital component is missing in the current modelling and the corresponding analysis of RCP-OTFS, which 
%\del{instead} 
assumes that the rectangular pulse-shaped RCP-OTFS signal is fully received without bandpass filtering at the receiver, but the noise is restricted to a \emph{narrowband} corresponding to the sampling rate. 

In this letter, we study the performance of the rectangular pulse based RCP-OTFS, when the  receiver is equipped with a practically necessary BPF.
% \red{Without being able to handle the OOBE of rectangular pulse-shaped RCP-OTFS, how RCP-OTFS performs when a practical receiver is employed is an interesting and important question. In this letter, we address this question and analyse the impact of rectangular pulse on RCP-OTFS performance.} 
% Specifically, a bandpass filter (BPF) at RF front-end of the OTFS receiver is taken into consideration. It is important to note that the BPF is a \emph{must} part for a practical receiver before all sampling and processing operations. Primarily, it ensures that no aliasing occurs so that the discrete waveform represents the underlying continuous waveform exactly. The BPF also restricts the band-unlimited additive white Gaussian noise (AWGN) to narrowband.
We show that with BPF, the transmitted DD domain symbols of rectangular pulse-shaped OTFS shift away from their initial constellation point and suffers from interference from other DD domain symbols. We shall point out that such interference is due to the rectangular pulse and BPF, which is different from the interference analysed in previous work such as [7], which is caused by the channel dispersion. Our further analysis shows that the bandpass filtered waveform experiences an unevenly distributed distortion across the DD grid, and demonstrates that the distortion introduces severe error floor for the OTFS error performance with a message passing detector.
\section{OTFS System model}
In an OTFS transmitter, symbols $X[k,l]$ from a complex signal constellation are mapped to a DD domain grid, with \(k\in{\{0,1,...,N-1\}}\) and \(l\in{\{0,1,...,M-1\}}\) representing their Doppler and delay indices, respectively. The signal is then converted to TF domain by applying an ISFFT \cite{Hadaniwcnc17}
\begin{equation}\label{isfft}
\mathcal X[n,m] = \frac{1}{\sqrt{MN}}\sum_{k=0}^{N-1} \sum_{l=0}^{M-1} X[k,l]e^{j2\pi(\frac{nk}{N}-\frac{ml}{M})},
\end{equation}
where \(m\in{\{0,1,...,M-1\}}\) indicates the frequency index and \(n\in{\{0,1,...,N-1\}}\) represents the time index. The TF domain form of the signal is equivalent to \(N\) OFDM symbols with \(M\) subcarriers. Orthogonality is achieved when the subcarrier spacing  \(\Delta f\) and OFDM symbol duration \(T\) adheres to the relationship \(\Delta f=\frac{1}{T}\). Next, the Heisenberg transform is performed to yield the time domain signal\cite{Hadaniwcnc17}
\begin{equation}\label{st}
s(t)=\sum _{n=0}^{N-1} \sum _{m=0}^{M-1}\mathcal X[n,m]e^{j2\pi m \Delta {f} (t-nT)} g_{\mathrm {tx}}(t-{nT}),
\end{equation}
where $g_{\mathrm {tx}}(t)$ defines the OFDM symbol-wise pulse shape at transmitter. Frame-wise CP of length \(L\)-sample is added, corresponding to extending the time {span} of the signal to \([-L\frac{T}{M},NT]\), such that \(L\frac{T}{M}\) is longer than the estimated channel delay spread~\cite{hadani2018arxiv}. 

When the signal is passed through a time-frequency dispersion channel, the received signal is given by
\begin{equation} r(t) = \int {\int a(\tau,\nu)e^{j2\pi \nu (t - \tau)}s(t - \tau) d\tau} d\nu + w(t),
\end{equation}
where \(a(\tau,\nu)\) denotes the spreading function of the channel and \(w(t)\) denotes AWGN. It is further assumed that the equivalent sampling channel, representing the physical channel subject to the delay-Doppler resolution \(\{\frac{T}{M},\frac{1}{NT}\}\), has $P$ paths on the DD grid, which leads to \cite{bello}
\begin{equation} a(\tau,\nu) = \sum _{i=1}^{P} a_{i} \delta (\tau -\tau _{i}) \delta (\nu -\nu _{i}), \end{equation}
where \(\delta\) denotes Dirac Delta, \(a_{i}\) is the path gain, \(\tau _{i}\) is the delay as an integer multiple of \(\frac {1}{M\Delta f}\), and \(\nu _{i}\) is the Doppler as an integer multiple of \(\frac {1}{NT}\) of the \(i\)th path.

The receiver performs Wigner transform on \(r (t)\) followed by sampling, yielding the discrete TF domain signal [2].
\begin{equation} 
\mathcal Y[n,m] = \int g_{\mathrm{ rx}}^{*}(t'-t) r(t') e^{-j 2 \pi f (t'-t)} dt'|_{t=nT,f=m\Delta {f}}, 
\end{equation}
where \( g_{\mathrm {rx}}(t)\) is the receiver pulse. For rectangular pulse shape,
\begin{equation*}
    g_{\mathrm{ tx}}(t) = g_{\mathrm{ rx}}(t) = \begin{cases} 1 & 0\leq t\leq T, \\
    0 & otherwise.
    \end{cases}
\end{equation*}
Note that the process in (5) is equivalent to first sampling in time domain and doing all transforms in a discrete form, as is performed by a practical receiver. Finally, the TF domain received signal is transformed to DD domain via simplectic finite fourier transform (SFFT), 
\begin{equation*} 
    Y[k,l] = \frac{1}{\sqrt{NM}} \sum _{n=0}^{N-1} \sum _{m=0}^{M-1} \mathcal{Y}[n,m] e^{-j2\pi \left({\frac{nk}{N}-\frac{ml}{M}}\right)},
\end{equation*}
where detection is performed.
\begin{figure}[!t]
\centering
\includegraphics[width=3.3in]{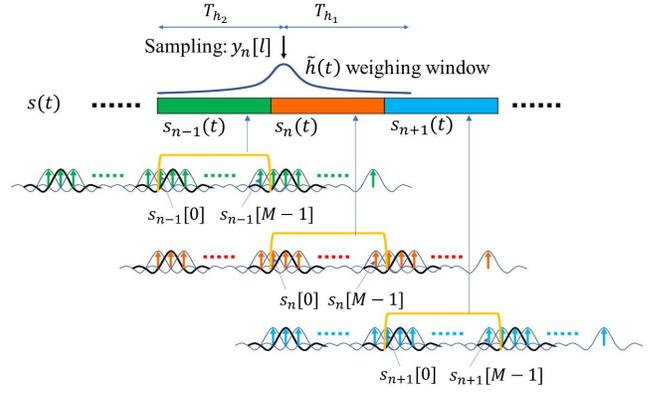}
\caption{The equivalent process of Heisenberg Transform and effect of filtering on OTFS signals. The yellow window indicates the per-OFDM pulse \(g_{\mathrm{ tx}}(t)\). \(\Tilde{h}(t)=h(-t)\) represents the mirror image of \(h(t)\) }
\label{fig_diag_heis}
\end{figure}

\section{Interference Analysis}\label{section:3}
\subsection{Interference in discrete time domain}
%For ease of analysis, we first rewrite the transmitted signal $s(t)$ in (\ref{st}) into a form which is interpolated and windowed from 
At the transmitter, the $MN$ discrete time domain samples of $N$ % $M$-subcarrier 
OFDM symbols in an OTFS frame can be given by 
\begin{equation}\label{discret_s}
s[nM+l]=\sum _{m=0}^{M-1} \mathcal X[n,m]e^{j\frac{2\pi m l}{M}},
\end{equation}
with $0\le n\le N-1$ and $0\le l\le M-1$. Then, for the $n$th OFDM symbol (the $n$th time slot), we have $M$ samples $\mathbf s_n=[s[nM],\ldots,s[nM+M-1]]$. Combining  the ISFFT in (\ref{isfft}) and the IDFT in (\ref{discret_s}),  the $l$th element of $\mathbf s_n$ can be obtained from the inverse discrete Zak transform as 
\begin{equation}
    s_n[l]=s[nM+l]=\frac{1}{\sqrt{N}}\sum_{k=0}^{N-1} X[k,l]e^{j\frac{2\pi nk}{N}}.
\end{equation}
For ease of analysis, we rewrite the transmitted signal $s(t)$ in (\ref{st}) into a form which is interpolated and windowed from $MN$ samples in (\ref{discret_s}). The process is also shown in the lower part of Fig.~\ref{fig_diag_heis}.
%The pulse shaping of the $n$th OFDM symbol can be performed by first interpolating the cyclically extended $\mathbf s_n$ namely $[\ldots,\mathbf s_n, \mathbf s_n, \mathbf s_n, \ldots]$  using the ideal interpolation filter $\sinc(t\frac{M}{T})$ 
% to generate $\tilde s_n(t)= \sum_{m=-M/2}^{M/2-1} \mathcal X[\big(m\big)_M,n] e^{j\frac{2\pi m}{T}t}$, where $\big(\cdot\big)_{M}$ denotes modulo $M$.
% Then, a transmit pulse $g_{tx}(t)e^{j\frac{\pi M}{T}t}$ based windowing is applied to $\tilde s_n(t)$ to obtain the analog OFDM symbol $s_n(t)$, where the subcarriers are shifted by the extra term of $e^{j\frac{\pi M}{T}t}$ to coincide with positive frequencies only in (\ref{st}).
Let \(\big(\cdot\big)_{M}\) denotes modulo \(M\), assuming a $T$-duration frame-wise CP,
%and no signal presents after the frame terminates, 
the OTFS frame can be written as
%it has a slot-wise representation given by
\begin{equation*}
    s(t) = \sum_{n=-1}^{N}s_{n}(t)
\end{equation*}
where the $n$th $T$-duration slot is 
\begin{equation}\label{Heis_equiv}
    s_n(t)=g_{\mathrm {tx}}(t-nT)\sum_{l'=-\infty}^{\infty}s_n[\left(l'\right)_{M}] \sinc\left(t-l'\frac{T}{M}-nT\right), 
\end{equation}
for $0\le n \le N-1$, while $s_{-1}(t)=s_{N-1}(t+NT)$ and $s_N(t)=0$ correspond to the frame-wise CP and the frame termination, respectively. Obviously, because of $g_{\mathrm {tx}}(t)=0$ for $t\not\in \left[0,T\right]$, we have $s_n(t)=0$ for $t\not\in \left[nT,(n+1)T\right]$.
% \begin{equation}\label{H_equi} 
%   s_{n}(t) = \begin{cases}  \begin{aligned} g_{\mathrm {tx}}(t-nT)\sum _{l'=-\infty}^{\infty}s_{n}[\left(l'\right)_{M}] \\
%   \times \sinc\left(t-l'\frac{T}{M}-nT\right) 
%   \end{aligned} & \begin{aligned}
%   n = 0,1,\\
%   \dots,N-1    
%   \end{aligned} \\
%   s_{N-1}(t) & n=-1\\
%   0 & n = N
%   \end{cases}
% \end{equation}
% within which \(\big(\cdot\big)_{M}\) denotes modulo \(M\), and 
% \begin{equation}\label{Zak}
%     s_{n}[l] = \begin{cases}
%     \sum_{k=0}^{N-1} x[k,l]e^{\frac{j2\pi nk}{N}} &  n = -1,0,\dots,N-1 \\
%     0 & n = N
%     \end{cases}
% \end{equation}
% \begin{equation*}
%     \sinc'(t) = \frac{\sin(\pi t)}{\pi t}e^{j\pi t}
% \end{equation*}
% In (\ref{Zak}), \(s_{n}[l]\) denotes the IDFT of \(X[m,n]\) along frequency direction for a given \(n\), but accommodated to the new range of \(n\), which could also be obtained from the discrete Zak transform. 

%

The main focus of the letter is to analyse the rectangular pulse's effect on OTFS's receiver performance. To provide some insight and simplify tedious analysis, 
%we consider an ideal channel \( a(\tau,\nu) = \delta (\tau ) \delta (\nu) \)~\cite{viterbo_twc_18} and temporarily ignore AWGN.
we consider the AWGN channel with a spread function \(a(\tau,\nu) = \delta (\tau ) \delta (\nu) \) and temporarily ignore the noise.
In that case, the received signal at the RF front-end is identical to the transmitted signal, namely, \(r(t) = s(t)\) except that a BPF is applied to \(r(t)\) before sampling. %Note that this ideal channel presents that the receiver performance is solely affected by the rectangular pulse and not by the channel. 
Considering the AWGN channel means that the receiver performance is solely affected by the rectangular pulse {with BPF} and not by the channel dispersion.
In Section IV, we will also show the OTFS performance over practical doubly-selective channels.
%, which shows that this analysis provides an error performance lower bound for practical channels. }

%In our analysis, we consider the ideal channel \( a(\tau,\nu) = \delta (\tau ) \delta (\nu) \)~\cite{viterbo_twc_18} and temporarily ignore AWGN. \add{While practical channel for OTFS use cases is doubly-dispersive, analysis can be significantly complicated and distract the focus from the rectangular pulse's effect on receiver performance, which the observations under ideal channel will suffice to illustrate.} Note that considering this ideal channel represents the best case or the least effect from the rectangular pulse compared to practical channels. 
%\\The received signal at RF front-end of receiver is then identical to the transmitted signal, namely, \(r(t) = s(t)\) except that a BPF is applied to \(r(t)\) before sampling. \add{Hence, in the rest of this section, any distortion of OTFS signal is solely caused by the rectangular pulse shape.}

Now we consider the BPF of the practical receiver. Ideally, the bandwidth of this BPF should be just enough to pass the signal, while rejecting as much noise as possible from the band where spectral component of the signal is insignificant. We assume that the lowpass equivalent of the BPF, referred to as LPF from hereon, is a non-causal filter with its impulse response \(h(t)\) supported on \(\left[-T_{h1},T_{h2}\right]\) and peaks at \(t=0\). This ensures that equivalent sampling at receiver could be made at integer multiples of \(\frac{T}{M}\), and change of the LPF causality only introduces a group delay that can be equalized easily. Then, let $\mathbf y$ denote the discrete time samples of the filtered receiver signal and $\mathbf y_n=[y[nM],\ldots,y[nM+M-1]]$ denote the samples in the \(n\)th time slot of $\mathbf y$, we have
% _{-T_{h2}+nT}^{T_{h1}+nT} & = \int{h\left(nT+l\frac{T}{M}-t'\right) s(t')dt'}, 
\begin{align}\label{t_conv}
    y_{n}[l]\coloneqq y[nM+l] & =  r(t) * h(t)|_{t=nT+l\frac{T}{M}} \nonumber\\ 
    & = \int{h\left(nT+l\frac{T}{M}-t'\right) s(t')dt'},  
    %& =  \int_{-T_{h1}}^{T_{h2}} s\left(nT+l\frac{T}{M}-t'\right) h(t')dt', 
\end{align}
where $0\le n \le N-1$, $0\le l \le M-1$, and $*$ denotes convolution. 
%{It simply performs a weighted integration of \(s(t)\) over the support of the mirror image of \(h(t)\). denoted by \(\Tilde{h}(t)\)} \red{not sure what is the purpose of this?} 
The effect of filtering on the OTFS signal is shown in Fig. \ref{fig_diag_heis}. It can be seen that the integration in (\ref{t_conv}) includes signals spanned by the discrete samples in the preceding, current and succeeding slots, which introduces interference. % \red{between signals from different time slots}.
%\uline{Furthermore, the rectangular window used at transmitter also destroys the orthogonality between time domain pulses and causes interference between symbols within the slot, which will be illustrated next.}

In the following interference analysis, we consider the practical case where the impulse response of the LPF is shorter than the slot length, namely, \(T_{h1},T_{h2} <T\). 
%Given that \(g_{\mathrm {tx}}(t)\) is only supported on \(t\in \left[0,T\right]\) and from considering (\ref{Heis_equiv}), (\ref{t_conv}) can be simplified as follows,
Recall that $s_n(t)=0$ for $t\not\in \left[nT,(n+1)T\right]$, (\ref{t_conv}) can be simplified as
\begin{align} \nonumber %\label{rx_mix} 
    y_{n}[l] = \sum_{q=-1}^{1}\underbrace{\int_{ (n+q)T}^{(n+q+1)T}{h\left(nT+l\frac{T}{M}-t'\right)s_{n+q}(t')dt'}}_{y_{n}^{(q)}[l]},    \nonumber 
\end{align}
% \begin{align}
%     r_{n}[l] &=\sum_{\dot n=n-1}^{n+1}{\int_{\dot nT}^{(\dot n+1)T}{h\left(nT+l\frac{T}{M}-t'\right)s_{\dot n}(t')dt'}} \nonumber\\
%     & = \sum_{q=-1}^{1}r_{n}^{q}[l]    
% \end{align} 
% \begin{align}\label{rx_mix}
%     r_{n}[l] &=\sum_{q=-1}^{1}\int_{ (n+q)T}^{(n+q+1)T}{h\left(nT+l\frac{T}{M}-t'\right)s_{n+q}(t')dt'} \nonumber\\
%     & = \sum_{q=-1}^{1}r_{n}^{q}[l],     
% \end{align}
%= \sum_{\substack{q=-1,0,1}}r_{n}^{q}[l]  
%\begin{align}
    %r_{n}[l] &=\int_{(n-1)T}^{(n+1)T}{h\left(nT+l\frac{T}{M}-\tau\right)s(\tau)d\tau} \nonumber\\
    %& = \sum_{\substack{q=-1,0,1}}r_{n}^{q}[l]    
%\end{align}
where \(y_{n}^{(q)}[l]\) denotes the signal received at the \(l\)th delay position in the \(n\)th time slot due to the contribution from the preceding, current and succeeding slots when \(q=-1,0,1\), respectively. %(\ref{rx_sep}) and  (\ref{int_per_d}) can then be obtained by \(s_{n+q}(t')\) in
Substituting  (\ref{Heis_equiv}) into \(y_n^{(q)}[l]\),  we obtain % and change variable \(t'\leftarrow t'-nT\), \(r_{n}[l]\) can be rewritten
\begin{align}\label{rx_sep}
    y_{n}^{(q)}[l] = & \sum _{l'=-\infty}^{\infty}s_{n+q}[\big(l'\big)_{M}]\times \nonumber\\
     & \quad \underbrace{\int_{qT}^{(q+1)T}h\left(l\frac{T}{M}-t'\right)\sinc{\left(t'-l'\frac{T}{M}\right)}dt'}_{i^{(q)}[l,l']},   
\end{align}
where
$i^{(q)}[l,l']$
% \begin{equation}\label{int_per_d}
%     i^{q}[l,l'] = \int_{qT}^{(q+1)T}h\left(l\frac{T}{M}-t'\right)\sinc{\left(t'-l'\frac{T}{M}\right)}dt' 
% \end{equation}
represents the level of interference caused by the \(l'\)th discrete time symbol from the preceding, current and succeeding slots on the \(l\)th discrete time symbol in the current slot, for \(q=-1,0,1\), respectively. 
%\att{Here, all three quantities are only dependent on \(h(t)\), \(l\) and \(l'\). }

If \(s(t)\) is a strictly band-limited signal without OOBE, the receiver LPF would not change the signal and hence introduces no interference. In that case, we have the orthogonality between the sample-wise transmitting pulse and the LPF impulse response~\cite{RChang1966}
\begin{align}
     \int_{-\infty}^{\infty}{h\left(l\frac{T}{M}-t'\right)\sinc{\left(t'-l'\frac{T}{M}\right)}dt'}=\delta{\left[l-l'\right]},
\end{align}
which is analogous to Nyquist Criterion for pulses. Note that this orthogonality holds only when the entire support of sinc and \(h(t)\) is included in the integration. Unfortunately, the rectangular pulse of OTFS breaks the integrity of integration range and results in non-trivial values of $i^{(q)}[l,l']$.
%, or equivalently, the sample-wise pulse shape of the transmitted signal becomes a truncated sinc function by a \red{window?} that varies with delay index, which is no longer orthogonal to \(h(t)\). \red{so what? what do you want to say here? or leading the readers to you analysis in the following.}

\subsection{Interference analysis in DD domain}
The received signal and interference can be analysed in the DD domain by performing a discrete Zak Transform on the $MN$ time domain discrete samples from (\ref{t_conv}).  %{Therefore, to derive exact form of interference level in DD domain, it is equivalent to rearrange the discrete time samples of the received signal \(r_{n}[l]\) into time-delay representation and convert it to DD domain by performing $N$-point-DFT along the time dimension.} 
First, let us represent \(l'\) in (\ref{rx_sep}) as \(l'=l_{2}M+l_{1}\), with \(0\leq l_{1}\leq M-1\) and \(-\infty< l_{2}< \infty\). %\del{\del{Thus, the} \add{The} received \del{signal} \red{symbols} of each \del{time-slot} \add{OTFS signal} can be divided into $M$ groups, where each group \del{contains the interference caused by} \add{corresponds to} the discrete time symbols with the same delay index \(l_{1}\).}
For a given \(q\), we then have
\begin{align}
    y_{n}^{(q)}[l] & = \sum_{l_{1}=0}^{M-1}\sum _{l_{2}=-\infty}^{\infty}s_{n+q}[l_{1}]i^{(q)}[l,l_{2}M+l_{1}] \nonumber \\
    & = \sum_{l_{1}=0}^{M-1}s_{n+q}[l_{1}]\underbrace{\sum _{l_{2}=-\infty}^{\infty}i^{(q)}[l,l_{2}M+l_{1}]}_{I^{(q)}[l,l_{1}]}, 
\end{align}
where $ I^{(q)}[l, l_{1}] $ represents the level of interference caused by the \(l_1\)-th discrete time symbol from the preceding \((q=-1)\), current \((q=0)\) and succeeding \((q=1)\) slots on the \(l\)th discrete time symbol in the current slot, respectively,  where $l \in \{0,1,\dots, M-1\}$ and $l_{1}\in \{0,1,\dots, M-1\}$.
%Note that when \(l_{1}=l\) and \(q=0\), the particular summation component of \(r_{n}^{q}[l]\) is the desired signal \(s_{n}[l]\) scaled with \(I^{q}[l,l]\). This leads to a further decomposition of \(r_{n}^{q}[l]\), given by
From hereon, we use the time delay representation for the transmitted and receiver signals as \(s[n,l]=s_{n}[l]\) and \(y[n,l]=y_{n}[l]\), where $0\le n \le N-1$ denotes the time index and $0\le l \le M-1$ denotes the delay index. The DD domain discrete samples can thus be given by
%\begin{align}\label{eq_dd_received}
    %&Y[k,l] = \frac{1}{\sqrt{N}}\sum_{k'=0}^{N-1}y[n,l]e^{-j2\pi \frac{nk'}{N}}\nonumber \\
    %&= \sum_{\substack{l_{1}=0}}^{M-1}\underbrace{\left(\sum_{q=-1}^{1}I^{(q)}[l,l_{1}] \left\{\frac{1}{\sqrt{N}}\sum_{k'=0}^{N-1} \left( s[n+q,l_{1}]\right) e^{-j2\pi \frac{nk'}{N}}\right\} \right)}_{Y'[k,l,l_{1}]} 
%\end{align}
\begin{align}\label{eq_dd_received}
    &Y[k,l] = \frac{1}{\sqrt{N}}\sum_{n=0}^{N-1}y[n,l]e^{-j2\pi \frac{nk}{N}} \\
    &= \sum_{\substack{l_{1}=0}}^{M-1}{\left(\sum_{q=-1}^{1}I^{(q)}[l,l_{1}] \left\{\frac{1}{\sqrt{N}}\sum_{n=0}^{N-1} \left( s[n+q,l_{1}]\right) e^{-j2\pi \frac{nk}{N}}\right\} \right)}.\nonumber
\end{align}
By (5), for any delay index \(l_{1}\), the DFT of \(s[n,l_{1}]\) yields exactly \(X[k,l_{1}]\), while
$s[n-1,l_{1}] = s[\big(n-1\big)_{N},l_{1}]$ and $s[n+1,l_{1}] = s[\big(n+1\big)_{N},l_{1}] + s_{c}[n,l_{1}]$
% \begin{align*}
%     s[n-1,l_{1}] &= s[\big(n-1\big)_{N},l_{1}] \\
%     s[n+1,l_{1}] &= s[\big(n+1\big)_{N},l_{1}] + s_{c}[n,l_{1}]
% \end{align*}
for \(n = 0,1,\dots,N-1\), where
\begin{equation*}
   s_{c}[n,l_{1}] = \begin{cases} 0 & n = 0,1,\dots,N-2 \\
    -\frac{1}{\sqrt{N}}\sum_{k=0}^{N-1} X[k,l_{1}] & n = N-1
    \end{cases} 
\end{equation*}
Now let us analyse (\ref{eq_dd_received}). % and (\ref{eq_I_delay_I}).  
Since a cyclic shift in time domain corresponds to linear phase variation in discrete Doppler domain, for any \(l_{1}\), we have
%\begin{align}\label{eq_dd_form}
     %Y'[k,l,l_{1}] & = \sum_{\substack{q=-1}}^{1}{I^{(q)}[l,l_{1}]X[k,l_{1}]e^{\frac{j2\pi k}{N}q}} \nonumber\\ & \quad -  I^{(1)}[l,l_{1}]\bigg(\frac{1}{N}\sum_{k'=0}^{N-1}X[k',l_{1}]\bigg)e^{-\frac{j2\pi k(N-1)}{N}}
%\end{align}
\begin{align}\label{eq_dd_form}
     Y[k,l] & = \sum_{l_{1}=0}^{M-1}\Bigg(\sum_{\substack{q=-1}}^{1}{I^{(q)}[l,l_{1}]X[k,l_{1}]e^{\frac{j2\pi k}{N}q}} \nonumber\\ & \quad -  I^{(1)}[l,l_{1}]\bigg(\frac{1}{N}\sum_{k'=0}^{N-1}X[k',l_{1}]\bigg)e^{-\frac{j2\pi k(N-1)}{N}}\Bigg)
\end{align}
and hence,
\begin{equation}\label{eq_dd_y}
    Y[k,l] = C[k,l,l]X[k,l] + \underbrace{\sum_{(k',l_{1})\neq (k,l)}C[k',l,l_{1}]X[k',l_{1}]}_{I[k,l]}
\end{equation}
for \(k'\in{\{0,1,...,N-1\}}\) and \(l_{1}\in{\{0,1,...,M-1\}}\), where 
\begin{align}
& C[k',l,l_{1}] =   \nonumber \\ 
   & \quad\quad
   \begin{cases}
    \begin{aligned}
    &I^{(0)}[l,l_{1}]+I^{(-1)}[l,l_{1}]e^{\frac{-j2\pi k}{N}} \\ 
    & \quad + I^{(1)}[l,l_{1}]\left(e^{\frac{j2\pi k}{N}}-\frac{1}{N}e^{-\frac{j2\pi k(N-1)}{N}}\right)
    \end{aligned}, & k'=k \\
    -\frac{1}{N}I^{(1)}[l,l_{1}]e^{-\frac{j2\pi k'(N-1)}{N}}, & k'\neq k
    \end{cases} \label{eq_deviation}
\end{align}
The first term in (\ref{eq_dd_y}) denotes the desired signal received on the \([k,l]\)-th DD grid, scaled by a complex coefficient \(C[k,l,l]\), which represents the amplitude change and phase rotation for the transmitted symbol on the same grid.
On the other hand, the second term of (\ref{eq_dd_y}) stands for interference from symbols transmitted on all the grid points in DD domain with \((k',l_{1})\neq (k,l)\). Hence, we define \(I[k,l]\) as inter-delay-Doppler-interference (IDDI) for the symbol on the \([k,l]\)-th DD grid.
Note that it is a linear combination of \emph{i.i.d} variables \(X[k',l_{1}]\) with zero mean and variance \(\sigma_{s}^{2}\), where \(\sigma_{s}^{2}\) is the average symbol energy of \(X[k,l]\). Thus, the average IDDI power is given by
\begin{align}
   &V[k,l]=\text{E}\left\{I[k,l]\cdot I^{*}[k,l]\right\} \nonumber  \\
    &= \text{E}\left\{X[k,l]\cdot X^{*}[k,l]\right\} \sum_{(k',l_{1})\neq (k,l)}C[k',l,l_{1}]C^{*}[k',l,l_{1}]  \nonumber \\ 
    % & = \sigma_{s}^{2}\sum_{\substack{l_{1}\neq l,\\l_{1}=0}}^{M-1}\Big|C[k,l,l_{1}]\Big|^{2}+\frac{\sigma_{s}^{2}}{N^{2}}\sum_{l_{1}=0}^{M-1}\sum_{\substack{k'=0,\\k'\neq k}}^{N-1}\Big(I^{(1)}[l,l_{1}]\Big)^{2} \nonumber \\ 
    & = \sigma_{s}^{2}\sum_{\substack{l_{1}\neq l,\\l_{1}=0}}^{M-1}\Big|C[k,l,l_{1}]\Big|^{2}+\frac{\sigma_{s}^{2}(N-1)}{N^{2}}\sum_{l_{1}=0}^{M-1}\Big(I^{(1)}[l,l_{1}]\Big)^{2}, \label{eq_iddi_power} 
\end{align}
which shows its relation to \(k\), \(l\) and \(h(t)\). The signal to interference power ratio (SIR) on each \((k,l)\) DD grid, without considering the interference caused by the channel dispersion, can thus be quantitatively calculated as
\begin{equation}\label{eq_sir}
    \text{SIR}[k,l] = \frac{\sigma_{s}^{2}\Big|C[k,l,l]\Big|^{2}}{V[k,l]}.
\end{equation}

\section{Interference and performance Evaluation}
%In this section, the derivation of expected interference power from Section~\ref{section:3} is verified and compared with simulation, and error performance of the rectangular pulse shaped RCP-OTFS is {evaluated on both AWGN and doubly-dispersion channels}. In our simulation, the lowpass filter is a near-rectangular filter with a bandwidth \(\frac{M}{T}\) and its time response is a truncated sinc which centres at \(t=0\) and \(T_{h_{1}}=T_{h_{2}}=T\). Simulation parameters are summarized in \ref{table_para}. For practical channels, we consider Extended Vehicular A (EVA) model~\cite{EVA} for the delay of each path, and their associated Doppler are generated randomly with Jake's formula. %The DD domain signal has delay and Doppler dimension \(M=128\) and \(N=16\) respectively, and information bits are carried on 4-QAM symbols with unity power.

In this section, the derivation of expected interference power from Section~\ref{section:3} is verified and compared with simulation, and error performance of the rectangular pulse shaped RCP-OTFS is {evaluated for both the AWGN and doubly-selective channels}. In our simulation, the LPF is a near-rectangular filter with a bandwidth \(\frac{M}{T}\) and its time response is a truncated sinc which centres at \(t=0\) and \(T_{h_{1}}=T_{h_{2}}=T\). Other types of LPFs, such as Bessel, Chebyshev and Butterworth filters, etc., will have similar effect.

% \begin{comment}

% \begin{table}[!t]\label{table_para}
% \caption{Summary of Simulation parameters\label{tab:table1}}
% \centering
% \begin{tabular}{|c|c|c|}
% \hline
% %Parameter & Value & Value \\
% %\hline
% Channel Type & Ideal (AWGN) & EVA \\
% \hline
% Baseband Modulation & 4-QAM & 4-QAM \\
% \hline
% Number of subcarriers (M) & 128,512 & 512 \\
% \hline
% N & 16,32 & 32\\
% \hline
% Subcarrier spacing & 15kHz & 15kHz \\
% \hline
% Carrier frequency & 5GHz & 5GHz \\
% \hline
% Maximum UE speed & -- & 120km/h \\
% \hline
% \end{tabular}
% \end{table}
% \end{comment}

\begin{figure}
\centering
\includegraphics[width=3.3in]{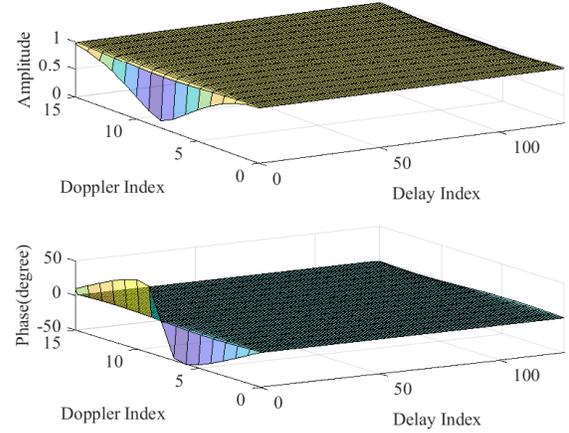}
\caption{Amplitude and phase variation for the transmitted DD domain signals}
\label{fig_shift_vector_new}
\end{figure}

% Removed on 20220724
%\begin{figure}
%\centering
%%\includegraphics[width=3.3in, height=1.8in]{z_int_level_4.eps}
%\includegraphics[width=3in]{z_int_level_final.eps}
%\caption{Estimated total interference power}
%\label{fig_av_int}
%\end{figure}

% \begin{comment}
% \begin{figure}
% \centering
% %\includegraphics[width=3.3in, height=1.5in]{z_sir_dd_4.png}
% \includegraphics[width=3.3in]{z_int_sir_comb.eps}
% \caption{SIR on delay-Doppler domain grid}
% \label{fig_sir_dd_1}
% \end{figure}
% \end{comment}

In Figs. 2 to 4, we demonstrate the effect of rectangular pulse and filtering by showing the amplitude and phase changes of signal component and interference on different DD grids in the AWGN channel. We use \(M=128\) and \(N=16\) as an example, which provides better visibility of interfered region in the figures. Fig.~\ref{fig_shift_vector_new} shows the amplitude change and phase rotation experienced by the transmitted DD domain symbols, described by \(C[k,l,l]\) in \eqref{eq_deviation}. For most symbols, \(C[k,l,l]\) is approximately a unity-length, zero-phase vector, which corresponding to almost no shift of the transmitted signal. Nonetheless, attenuation in power can be observed for symbols with delay index close to zero or \(M\) while Doppler index \(k\) is close to \(\frac{N}{2}\). This is due to relatively large values of \(I^{(-1)}[l,l]\) when \(l\) approaches zero and large values of \(I^{(1)}[l,l]\) when \(l\) approaches \(M\). When \(k\) is around \(\frac{N}{2}\), \(I^{(-1)}[l,l]e^{\frac{-j2\pi k}{N}}\) and \(I^{(1)}[l,l]e^{\frac{j2\pi k}{N}}\) are in opposite direction to the desired signal in the signal space, resulting in significant decrease of the original signal power.

%In Fig. 3, the average total interference power {over different Doppler} \add{indices} on each delay index is plotted. {It can be seen that} the estimated interference power fully agrees with the measurement from simulation, which justifies the derivations of {(\ref{IDoI_power}) and (\ref{IDI_power})}. We should point out that the inter-delay interference power is approximately $40$ dB larger than inter-Doppler interference (not shown here) and it contributes to the majority of total interference.

\begin{figure}
\centering
\includegraphics[width=3.3in]{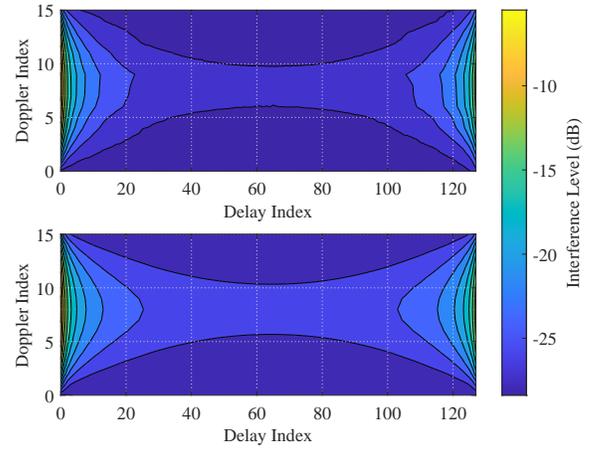}
\caption{Average interference power on delay-Doppler domain grid: (a) upper sub-figure: simulation result (b) lower sub-figure: derivation results from (17)}
\label{fig_int_dd}
\end{figure}

\begin{figure}
\centering
\includegraphics[width=3in]{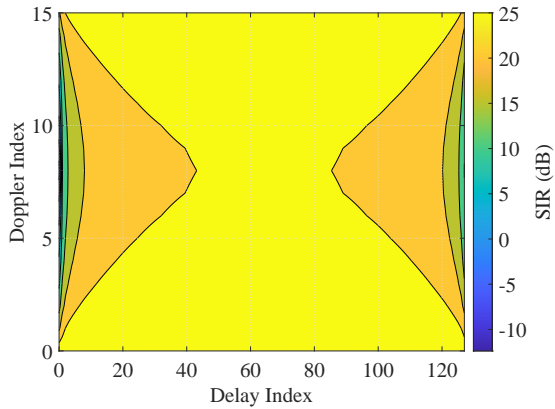}
\caption{SIR on delay-Doppler domain grid}
\label{fig_sir_dd}
\end{figure}

A description of average IDDI power and its variation with delay and Doppler indices is shown in Fig.~\ref{fig_int_dd}. The upper sub-figure presents the average interference power measured from simulation and the lower one is a plot of derivation result in \eqref{eq_iddi_power}. {It can be seen that} the estimated IDDI power agrees with the measurement from simulation. The contours of interference level is of hyperbola-like shape. Severer interference will be experienced by a symbol when its delay index is closer to 0 or \(M-1\), and its Doppler index is closer to \(N/2\). The SIR on each delay-Doppler grid is shown in Fig.~\ref{fig_sir_dd}, which follows a similar shape. Note that SIR is below 0 dB with Doppler index \(k\) close to \(\frac{N}{2}\) and delay index \(l\) closer to 0, which will dominate system performance when SNR is high. 
%The least influenced region is at SIRs around $26$ dB, indicating that error performance of rectangular pulse-shaped OTFS will be limited at high SNR. %, even if the severely influenced region is avoided by employing unloading techniques to mitigate OOBE as in conventional OFDM systems.

\begin{figure}
\centering
\includegraphics[width=3.0in]{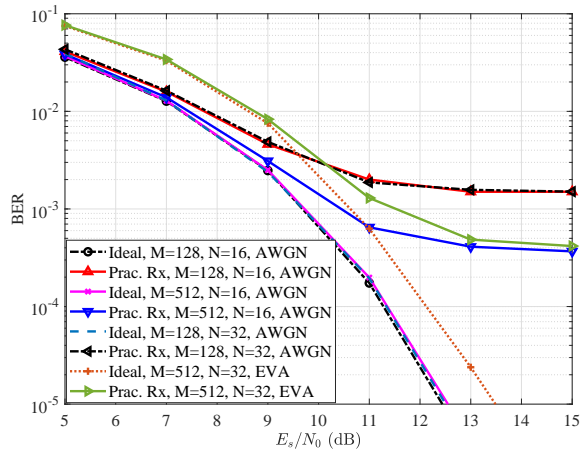}
\caption{BER performance of rectangular pulse-shaped RCP-OTFS. For the MP algorithm, damping factor is 0.7, and maximum iteration number is 50.}
\label{fig_err_perfo_1}
\end{figure}

The bit error rate (BER) performance of the OTFS system {in the AWGN and doubly-selective channels} is presented in Fig.~\ref{fig_err_perfo_1}, where \(E_{s}=\sigma_{s}^{2}\) and \(N_{0}\) is power spectral density of AWGN. DD domain 4-QAM symbols are employed to carry information. The message passing (MP) algorithm from \cite{viterbo_twc_18} is used for data detection. The nodes connection in the factor graph of the MP algorithm is governed by the input-output relationship for rectangular pulse-shaped OTFS given by Equation (24) in [7], which is widely accepted yet has not considered the effect of BPF.
%\add{For 'ideal' cases, we do not consider the BPF, but assume that the band-unlimited rectangular pulse-shaped RCP-OTFS can be fully received and critically sampled without aliasing, and AWGN is automatically restricted to a bandwidth of \(\frac{M}{T}\).}
Moreover, for the ``ideal" cases in Fig. 5, we assume that the theoretically band-unlimited rectangular pulse-shaped RCP-OTFS is fully received without BPF, while the AWGN is restricted to a bandwidth of \(\frac{M}{T}\). Note that such an ideal assumption is not realizable in practice.

In the AWGN channel, we can see from Fig. 5 that for the rectangular pulse-shaped OTFS system with (\(M=128,  N=16\)) and a practical receiver, an error floor occurs at BER around \(10^{-3}\) when \(E_{s}/N_{0}\ge 13\ \text{dB}\), while {there is no error floor} in the ideal case without BPF. {Simulation results with other frame sizes of (\(M=512,N=16\)) and (\(M=128,N=32\)) are also plotted for comparison.} It can be noticed that there is also an error floor but occurs at a lower BER when \(M\) increases. This is because of a relatively smaller proportion of interfered region in the entire frame for a larger $M$. Furthermore, we observe that no performance change occurs when \(N\) varies. For practical channels, we choose \(M=512\) and \(N=32\) and use the Extended Vehicular A (EVA) model~\cite{EVA} as the delay profile. Doppler for each path is generated randomly with Jakes' formula \cite{jakes} under the assumptions that \(\Delta f=15\ \text{kHz}\), carrier frequency \(f_{c}=5\ \text{GHz}\) and maximum relative speed \(v_{max}=120\ \text{km/h}\). Again, the BER curve for practical receiver touches the error floor at BER slightly below $10^{-3}$, which is not observed in the ideal case. Similar effect on error performance can also be observed for higher order modulations, which is not shown due to space limitation.

\section{Conclusion}

In this letter, we investigate the effect of rectangular pulse-shaping on error performance of OTFS modulation for practical receiver with a BPF. Assuming OOBE not handled, we first show that the rectangular pulse can introduce time domain interference for RCP-OTFS when band-limitation is applied. In DD domain, such interference corresponds to amplitude change and phase rotation of transmitted symbols as well as inter-delay-Doppler interference, the formula for which are derived. Our analysis shows that symbols mapped to the edges of delay and the middle of Doppler on the DD grid are most severely corrupted. Finally, error performance for such waveform is evaluated for both AWGN and doubly-selective channels, where an error floor can be observed at high SNRs. %\add{We conclude that OOBE may be a major issue for RCP-OTFS before its practical application} 
%\\OTFS does not provide an intuitive option for guarding in time and/or frequency domain, which can become a problem for its practical implementation.
%\\the OOB-induced dilemma between self-interference and noise.

\bibliographystyle{IEEEtran}
\bibliography{otfs}

\vfill

\end{document}